\documentclass[footinbib,aps,pra,reprint,superscriptaddress]{revtex4-1}

\usepackage{amsmath,amssymb,bbm,bm,graphicx,upgreek}

\usepackage[absolute]{textpos}
\usepackage{xcolor}

\begin{document}
\title{Intensity-Modulated Fiber-Optic Voltage Sensors for Power Distribution Systems}

\author{Joseph M. Lukens}
\email{lukensjm@ornl.gov}
\affiliation{Quantum Information Science Group, Computational Sciences and Engineering Division, Oak Ridge National Laboratory, Oak Ridge, Tennessee 37831, USA}

\author{Nicholas Lagakos}
\affiliation{SmartSenseCom, Vienna, Virginia 22180, USA}

\author{Victor Kaybulkin}
\affiliation{SmartSenseCom, Vienna, Virginia 22180, USA}

\author{Christopher J. Vizas}
\affiliation{SmartSenseCom, Vienna, Virginia 22180, USA}

\author{Daniel J. King}
\affiliation{Power and Energy Systems Group, Electrical and Electronics Systems Research Division, Oak Ridge National Laboratory, Oak Ridge, Tennessee 37831, USA}

\begin{abstract}
We design, test, and analyze fiber-optic voltage sensors based on optical reflection from a piezoelectric transducer. By controlling the physical dimensions of the device, we can tune the frequency of its natural resonance to achieve a desired sensitivity and bandwidth combination. In this work, we fully characterize sensors designed with a 2~kHz characteristic resonance, experimentally verifying a readily usable frequency range from approximately 10~Hz to 3~kHz. Spectral noise measurements indicate detectable voltage levels down to 300 mV rms at 60 Hz, along with a full-scale dynamic range of 60 dB, limited currently by the readout electronics, not the inherent performance of the transducer in the sensor. Additionally, we demonstrate a digital signal processing approach to equalize the measured frequency response, enabling accurate retrieval of short-pulse inputs. Our results suggest the value and applicability of intensity-modulated fiber-optic voltage sensors for measuring both steady-state waveforms and broadband transients which, coupled with the straightforward and compact design of the sensors, should make them effective tools in electric grid monitoring.
\end{abstract}

\maketitle
\section{Introduction}

Fiber-optic sensors have emerged as powerful instruments for monitoring a variety of physical phenomena~\cite{Udd2011, Krohn2014, Lee2003, Culshaw2008}. Their passive nature, high bandwidth, light weight, and immunity (in certain configurations) to electromagnetic interference have motivated the development of many commercial products, with devices including gyroscopes~\cite{Lefevre2014}, hydrophones~\cite{Dandridge2019, Cranch2003}, and electric current transducers~\cite{Ziegler2009} proving particularly popular. In applications focused on monitoring electromagnetic fields specifically, the most common approaches have relied on the Faraday and Pockels effects, which rotate an optical probe field's polarization state in proportion to the magnetic or electric field to be measured, respectively~\cite{Massey1975, Rashleigh1979, Bohnert2002, Bohnert2005}. While extremely sensitive to the fields of interest, these phase-based optical sensors are unfortunately also highly sensitive to temperature and birefringence drifts, requiring complex and expensive control systems to compensate for them. These electromagnetic sensors typically use solid state lasers optimized for fiber-optic communication systems which operate at high data rates (MHz or GHz). Below $\sim$500 Hz, these lasers generally experience $1/f$ noise which introduces increasing measurement error at low frequencies. In addition to the cost of solid state lasers, backreflected light must also be eliminated, since it significantly increases the internal noise of the laser system, thus placing tight demands on, e.g., optical connector interfaces. While the exceptional sensitivity of interferometric voltage or current sensors may justify their cost in some situations, they are well suited primarily to lab applications and industrial settings where the environment is well controlled, particularly the temperature. But for field applications and for frequencies lower than 1~kHz, such sensors are not very appropriate. 

In contrast, building on some of the basic ideas leveraged in early fiber-optic sensor designs~\cite{Kissinger1967}, intensity-based sensors function by delivering light to a transducer element and collecting a return signal in a second fiber (or fibers), where the returning optical power depends directly on the physical phenomena of interest. Through appropriate design of the transducer, many phenomena can be sensed in this method, such as temperature, strain, acoustic waves, static pressure, acceleration, and vibration~\cite{Krohn1987, Bucaro2005, Bucaro2013}. Recently, these techniques have been applied specifically to current and voltage sensors designed for power systems~\cite{Lagakos2017}. But their performance has yet to be analyzed in detail in the literature, creating a need for thorough characterization of this sensor type in the wider application space of power distribution.

In this article, we describe and test a complete fiber-optic sensing system for three-phase voltage measurements. Based on a piezoelectric transducer and a fiber-optic probe, this specific system design is found to enable high-sensitivity voltage measurements up to 3~kHz, with a strong mechanical resonance at 2~kHz. We obtain complete frequency responses for all three electrical phases, along with noise floor measurements, allowing us to estimate the sensitivity and dynamic range throughout the usable bandwidth. Finally, compensating the spectral response of the measured sensor output, we demonstrate an equalization approach for accurate reconstruction of impulses, making these sensors well suited for real-time, high-bandwidth voltage monitoring in demanding power distribution environments.

\section{Sensor Physics and Design}
\label{sec:design}
\begin{figure}[tb!]
\centering\includegraphics[width=3.15in]{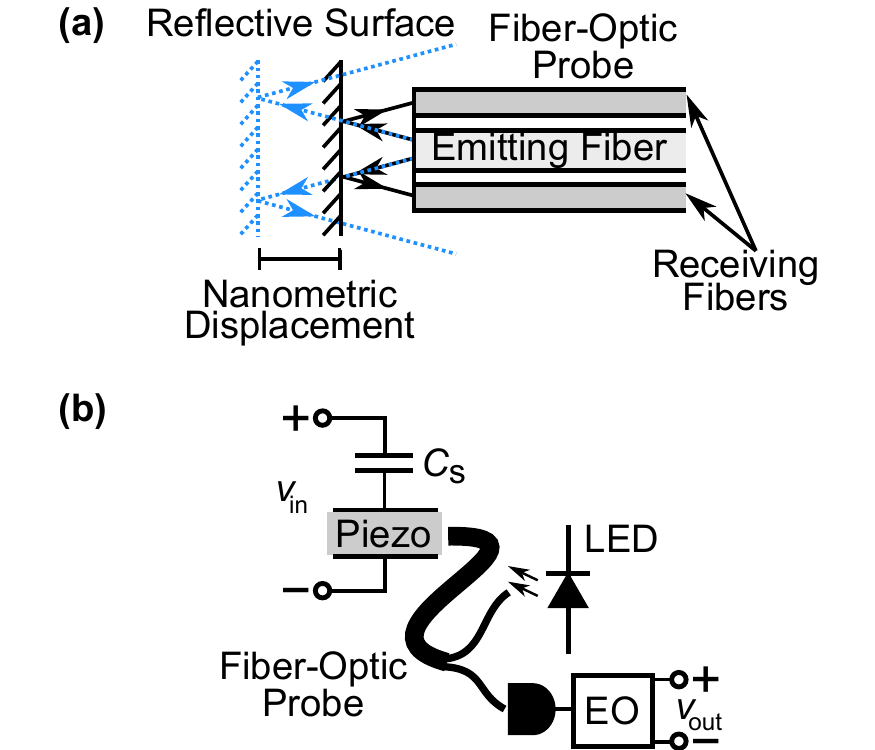}
\caption{Intensity-modulated fiber-optic sensor. (a) Basic principle of operation. Light is launched onto a reflective surface, the position of which is related to the quantity being sensed, and then collected by receiving fibers. (b) Voltage sensor configuration. A voltage applied across the series capacitor ($C_s$) and piezoelectric transducer (piezo) modulates the optical field that is collected by the return fibers, detected, and amplified with electro-optic circuitry (EO).}
\label{fig1}
\end{figure}

The fiber-optic intensity-modulated sensors considered here do not utilize electro- or magneto-optic effects but instead rely on a mechanical process, employing an optical probe to measure displacement of a transducer sensitive to the effect being measured. Both the sensitivity and responsivity depend on a variety of elements in transducer construction. Such sensors have extremely simple designs, require only moderate precision in fabrication and operation, demonstrate high linearity, and feature a wide dynamic range. A schematic of basic sensor operation is provided in Fig.~\ref{fig1}(a). Light in a central optical fiber propagates to the sensing element, where it is emitted from the fiber over a short distance (less than 1~mm) and bounces off of a highly reflective surface attached to an appropriate transducer. The reflected light is coupled into a fiber bundle symmetrically surrounding the emitting fiber which delivers the light to a photodetector converting optical power into an electrical signal. A force which displaces the sensing element modulates the distance between the end of the optical probe and the reflector, which in turn modulates the light power, as the amount of reflected light detected is a function of the displacement between the probe and the reflector. 

Now, because these intensity-modulated sensors rely on the mechanical displacement of a macroscopic-size reflective element, the transducer possesses natural resonance frequencies, in contrast to standard Pockels and Faraday effect phase-modulated sensors with no resonant features. In the particular system examined here, we utilize a specially designed bimorph transducer element constructed from PZT-4 piezoceramic (Navy Type I~\cite{DoD1995}) with nominal dimensions of $12\times 1.5\times0.5$~mm$^3$. This yields a fundamental cantilever resonance frequency of $\sim$2~kHz, which is confirmed by the tests in the following section. Importantly, because piezoelectric materials are so well understood and characterized, a variety of geometries could be considered, with dimensions chosen beforehand to realize a predefined resonance frequency. In general, there exists a design tradeoff between sensitivity and bandwidth: a lower fundamental resonance provides higher sensitivity, but reduced bandwidth, whereas upshifting the resonance frequency enables wider bandwidth at the cost of lower sensitivity. In this way, one can tailor the transducer to the bandwidth needs of the particular application.

Because of the straightforward, non-interferometric design of this intensity-based fiber-optic sensor, we employ LEDs as light sources which, when compared to lasers, are significantly less expensive, have a very long lifetime (mean time to failure measured in decades for the types of LEDs we utilize), and do not suffer as strongly from low-frequency intensity noise~\cite{Rumyantsev2004}. Because they do not require an optical cavity, LEDs are also less sensitive to backreflection-induced instabilities so that conventional fiber-optic connectors (and connection techniques) can be used. In order to optimize coupling efficiency, we utilize multimode fibers for transmission and collection as well. Figure~\ref{fig1}(b) depicts in simplified form the complete optical voltage sensor configuration for a single electrical phase, a commercial-grade sensor designed and constructed at SmartSenseCom. The specific fiber probe consists of seven identical multimode fibers each with a 200~$\upmu$m diameter glass core and 230~$\upmu$m plastic cladding, with a numerical aperture of 0.37. The single transmitting fiber is surrounded by six receiving fibers distributed in a fixed geometric pattern around the longitudinal axis~\cite{Bucaro2005}. Light from an LED emitting at 850~nm is coupled into the transmitting fiber and sent to the piezoelectric transducer, the surface of which is coated with a reflective film. Any voltage $v_\mathrm{in}$ applied across the capacitive voltage divider including the transducer stretches or compresses the material, modulating the optical power coupled into the return fibers. A photodiode then converts the received optical power into electric current, which is fed into an electro-optic (EO) circuit with a net transimpedance gain of $\sim5\times10^6$ V/A for amplification and filtering, yielding the output $v_\mathrm{out}$.

\begin{figure}[tb!]
\centering\includegraphics[width=3.15in]{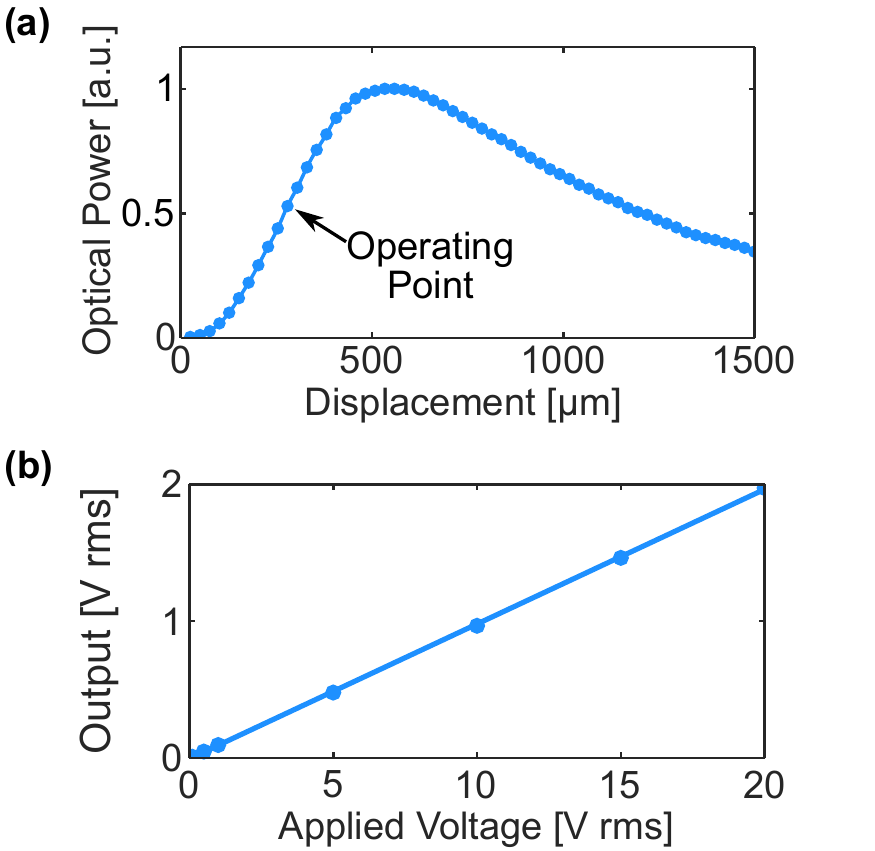}
\caption{Optical transducer characterization. (a) Collected optical power vs. fiber-probe/piezo separation. The chosen operating point at 280~$\upmu$m represents the highest slope. (b) EO readout potential at 280~$\upmu$m separation as a function of applied 100~Hz voltage. Near-ideal linear operation is obtained over these values (corresponding to peak-to-peak displacements up to $\sim$3~$\upmu$m).}
\label{fig2}
\end{figure}

To set the optimal sensor operating point, we first measure the collected optical power as a function of displacement between the fiber-optic probe tip and the piezo, controlled by a precision manual translation stage. An example curve is plotted in Fig.~\ref{fig2}(a). The reflected optical power initially increases with displacement until leveling off around 500~$\upmu$m and gradually decreasing thereafter. From these results, we set the quiescent displacement at 280~$\upmu$m, corresponding to the highest-slope region of the optical response. At this point, applying an oscillating voltage directly to the transducer terminals produces an extremely linear output from the electro-optic circuitry. Test results for a 100~Hz voltage applied to the piezo are shown in Fig.~\ref{fig2}(b); the log-log slope of the fit is $1.004 \pm 0.002$, extremely close to the ideal of unity expected for a perfect linear response.

In practice, the axial resolution of the optical probe is primarily limited by the noise of the readout circuitry. For the present EO configuration, this corresponds to a minimum detectable displacement on the order of 1.5~nm, or 0.05\% of our chosen maximum peak-to-peak displacement of 3~$\upmu$m (corresponding, at 100~Hz, to an applied potential of 20~V~rms), thereby giving a predicted dynamic range of $\sim$66~dB. Incidentally, as the maximum 3~$\upmu$m amplitude is barely visible on Fig.~\ref{fig2}(a), one should be able to increase the utilized displacement range several times over without significant loss of linearity, resulting in an even wider dynamic range and better resolution.

In order to set the overall scaling to a desired input voltage range, i.e., define the voltage-to-displacement gain factor, we place the transducer (which itself can be modeled as a capacitor) in a capacitive voltage divider circuit. Specifically, we choose the series capacitor $C_s$ in Fig.~\ref{fig1}(b) to scale the voltage applied to the piezo down by a factor of $\sim$10 compared to the input. We note that, for $v_\mathrm{in}$ under several kilovolts, this type of capacitive division works well. Alternatively, for significantly higher voltages (hundreds of kilovolts), a suitably scaled transducer potential can be achieved by using a simple antenna that diverts a \emph{de minimis} amount of energy from the electric field close to the active (hot) wire, and in this way provides electric-field--to--voltage conversion.

Finally, the components selected for this sensor system impart high stability with temperature. As this sensor relies on the principles of intensity modulation, it is unaffected by small temperature-induced variations in fiber path length, in contrast to phase-modulated--based sensing. Additionally, the chosen transducer ceramic, PZT-4, is extremely stable with temperature; its  $d_{31}$ piezoelectric coefficient varies by less than 1\% over the entire temperature range from $-$40 to $+$80 $^\circ$C~\cite{Hooker1998}. Empirically, we have not observed significant changes in sensor response from any other effects over  a range of $\sim$100 $^\circ$C, indicating this design (transducer and fiber-optic probe) should be well equipped for harsh field environments.

\section{Device Characterization}
\begin{figure}[tb!]
\centering\includegraphics[width=3.15in]{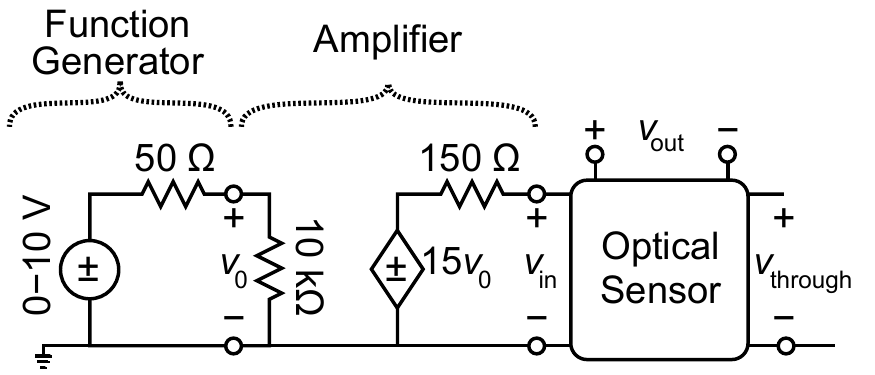}
\caption{Test setup for a single phase. A voltage amplifier driven by a function generator produces the optical sensor input $v_\mathrm{in}$. The feedthrough $v_\mathrm{through}$ is left open, and both $v_\mathrm{in}$ and the sensor output $v_\mathrm{out}$ are recorded on an oscilloscope. All impedances are nominal values from component datasheets.}
\label{fig3}
\end{figure}

For characterizing each voltage sensor as deployed within the complete test system, we adopt a black box view in terms of the input and output electrical connections, which include a three-phase plug input, three 0 to 5~V sensor outputs (one for each electrical phase), and a feedthrough three-phase output. We examine each phase separately in the test setup depicted schematically in Fig.~\ref{fig3}. A function generator (Stanford Research Systems DS345) drives a 10~kHz voltage amplifier (Thorlabs MDT694B) which produces the sensor test signal. Our particular amplifier is unipolar ($+$150~V~max), so that all waveforms have a 50\% dc offset; because the optical sensor detectors are capacitively coupled to reject low-frequency contributions ($<$10~Hz) and we still remain fully in the linear regime of the transducer displacement curve [Fig.~\ref{fig2}(b)], this offset has no observable effect on our test results. (Of course, dc sensing is possible by the transducer physics, but requires alternative detector amplifier and filter electronics.) We model the relationship between input potential $v_\mathrm{in}(t) = \tfrac{1}{2\pi} \int d\omega\, V_\mathrm{in}(\omega) e^{j\omega t}$ and sensor output $v_\mathrm{out}(t) = \tfrac{1}{2\pi} \int d\omega\, V_\mathrm{out}(\omega) e^{j\omega t}$ as a linear system with some (to be determined) frequency response $H(\omega)$, such that $V_\mathrm{out}(\omega)= H(\omega) V_\mathrm{in} (\omega)$, neglecting for the time being any nonlinear effects which would distort this behavior.

We then probe each phase with single-frequency sinewave excitations (phase to ground), so that the input signal can be modeled as $v_\mathrm{in}(t) = V_\mathrm{dc} + V_0 e^{j\omega t} + V_0^* e^{-j\omega t}$. Leaving the three-phase throughput open-circuited, we record the temporal waveforms $v_\mathrm{in} (t)$ and $v_\mathrm{out} (t)$ on a 200~MHz oscilloscope. By computing the fast Fourier transform (FFT) of both waveforms and taking the ratio of their values at the probe frequency $\omega_0$,  we retrieve the complex number $H(\omega_0)=V_\mathrm{out} (\omega_0 )/V_\mathrm{in} (\omega_0)$ which, by scanning $\omega_0$ through all values of interest, allows us to map out $H(\omega)$ completely. To reduce noise, we take the average of 16 traces, and we select a scope span setting as close as possible to 50 periods, with some variation due to the discrete horizontal division steps; for any given frequency the total number of recorded periods can never drop below 28, and the sampling rate is always at least 35 points per period.

\begin{figure}[tb!]
\centering\includegraphics[width=3.15in]{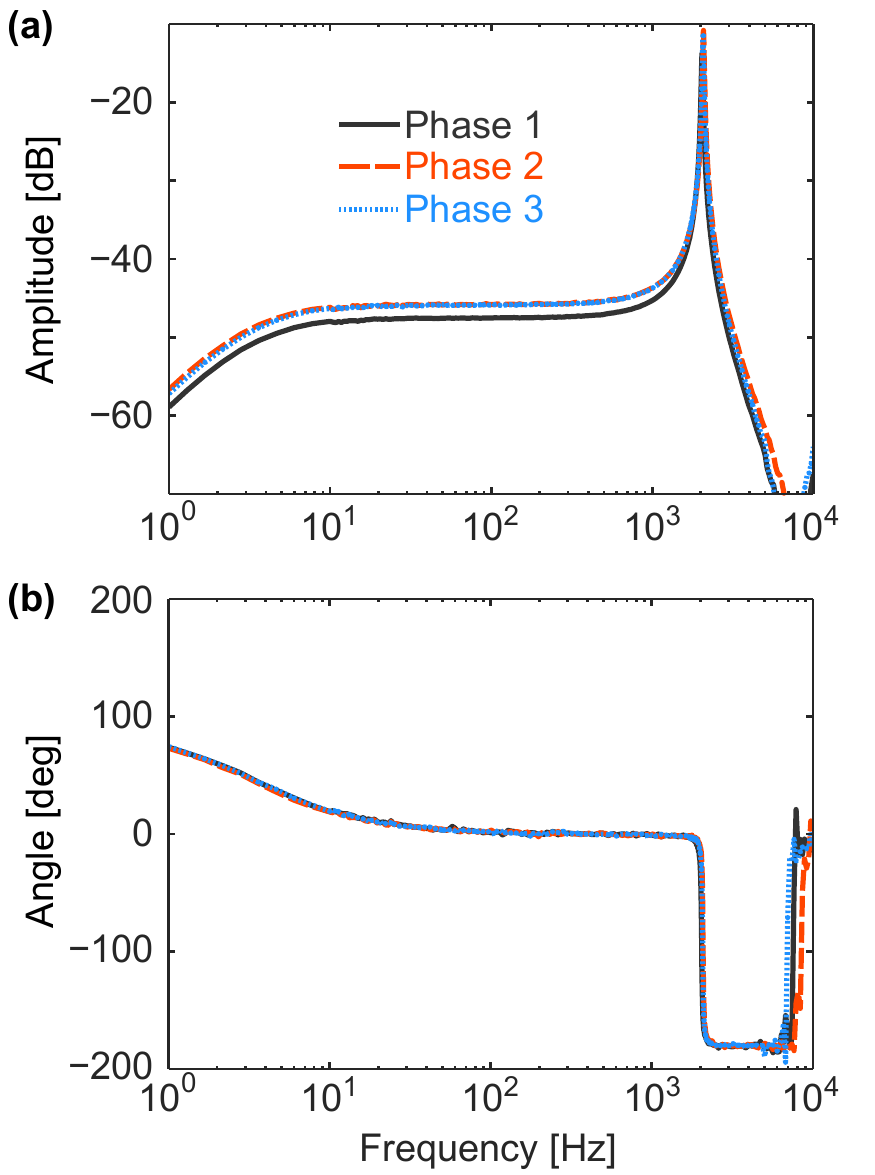}
\caption{Measured frequency responses of three-phase optical voltage sensors. (a) Amplitude $20\log_{10}|H(\omega)|$. (b) Phase angle $\angle H(\omega)$.}
\label{fig4}
\end{figure}

Bode plots of the frequency characterization results for all three sensors are presented in Fig.~\ref{fig4}. Apart from a slightly lower responsivity for the phase 1 sensor, all display extremely consistent behavior: an increase in amplitude until a very flat region from 10 to 500 Hz, followed by a strong resonance at 2~kHz and a steep rolloff thereafter. The $\sim$90$^\circ$ phase shift at low frequencies matches theory for a first-order high-pass filter (as does the amplitude's $\sim$20~dB/decade slope); and the sharp 180$^\circ$ drop at 2~kHz coincides with that of the universal resonance curve~\cite{Siebert1986}. If we define the effective bandwidth as comprising all frequencies with amplitudes equal to or above that of the flat region, these sensors permit useful monitoring from approximately 10 Hz to 3 kHz---or the first 50 harmonics of a 60~Hz voltage.

As noted above, the striking resonance observed here is a key distinguishing feature of this mechanical-transducer approach to voltage sensing. Incidentally, similar resonance effects have been reported in acoustic detectors based on intensity-modulated fiber-optic sensing principles~\cite{Bucaro2005}. From the perspective of voltage monitoring, the resonance simultaneously provides exceptional sensitivity for harmonics within a specific band but also introduces nuances which must be anticipated. For example, given the maximum sinewave output level of $\sim$1.4~V rms---set by the dc offset and 0~V minimum of the sensor output (cf. Fig.~\ref{fig6})---a 5~V rms input signal at 2.08~kHz would saturate the phase 2 EO output, whereas it would take $\sim$280~V~rms at 60~Hz to cause such saturation. Moreover, even in the absence of saturation, the spectral peak still reshapes the output signal so that it is not directly proportional to the input. Yet whereas nonlinear effects such as saturation prevent recovery of the original voltage, spectral distortion is \emph{linear} and can in principle be compensated digitally---a task we visit in Sec. \ref{sec:equal}. 

As with any sensing system, the sensitivity to input signals depends not only on the driven coherent response, but also on the system noise floor. In order to quantify this as well, we record the fluctuating output voltage $v_\mathrm{out} (t)$ with no signal applied to the input and estimate the periodogram, defined over the time interval $T$ as $S_T (\omega)= \tfrac{1}{T} \left| \int_0^T dt\,v_\mathrm{out} (t) e^{-j\omega t} \right|^2$, by computing the FFT of the raw samples. Repeating this process many times and averaging $S_T (\omega)$ at all frequencies returns an estimate of the true power spectral density $S(\omega)$~\cite{Papoulis2002}. To prevent aliasing in this broadband measurement, we precede the oscilloscope with a low-pass filter, using measurements with a 10~kHz lowpass filter (Thorlabs EF120) and 50~kS/s sampling to compute the noise level from 1 to 10~kHz, and a 1~kHz lowpass filter (Thorlabs EF110) and~5 kS/s to compute the noise from 10 to 1000~Hz with finer spectral resolution.

\begin{figure}[tb!]
\centering\includegraphics[width=3.15in]{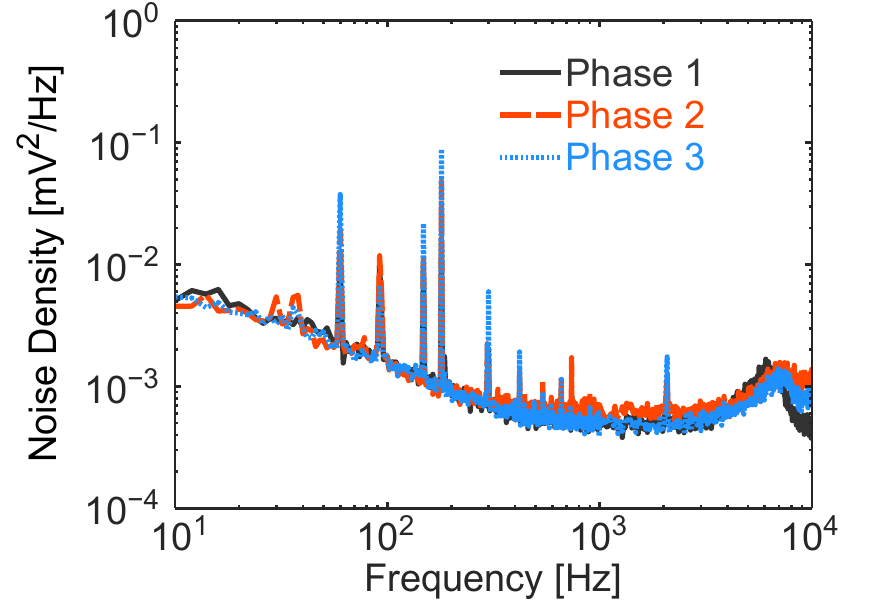}
\caption{Single-sided power spectral density of each sensor output, with no input signal applied.}
\label{fig5}
\end{figure}

The single-sided noise spectra [$S_+ (\omega)=2 S(\omega); \; \omega > 0$] for all three sensors are presented in Fig.~\ref{fig5}, where each point is the average of 128 separate periodograms. Spurs at 60~Hz and 2~kHz correspond to the wall plug ac power and natural sensor resonance, respectively. Similarly, the peaks at 180, 300, and 420~Hz are consistent with the known distortion characteristics of the power delivered to our laboratory building, whose third, fifth, and seventh harmonics have all been observed as particularly strong. The origin of spurs at 92 and 148~Hz is unknown and would require further study. Integrating over the sensors' full usable bandwidth (10 to 3000 Hz) and taking the root, we find rms noise values of 1.40, 1.54, and 1.48~mV for phases 1, 2, and 3, respectively. If we define the minimum detectable input signal as that which produces an rms output equal to the noise, the frequency responses of Fig.~\ref{fig4} indicate inputs as small as 300~mV~rms at 60 Hz and 5--7 mV rms at  resonance (depending on the phase) can be sensed. Coupled with the saturation limits discussed above, these sensors thus support a dynamic range of approximately 60~dB in the current configuration. This is a conservative estimate, in that it involves no additional narrowband filtering, which would significantly boost the sensitivity when, e.g., one is interested in monitoring only a specific frequency band. Importantly, the empirical dynamic range agrees reasonably well with the rough prediction of $\sim$66~dB based on the displacement sensitivity, as discussed in Sec.~\ref{sec:design}.

When operating in the field, significantly stronger noise spurs could be present, due to a variety of local environmental conditions. While analog filtering or shielding techniques could certainly be applied, the impact of spurs could also be removed digitally by incorporating, into the compensation technique described below, notch filters matched to the spur frequencies found from detailed \emph{in situ} spectral characterization at the particular installation. 

\section{Spectral Equalization}
\label{sec:equal}

With the frequency response $H(\omega)$ of all three sensors fully characterized, we now examine the important question of signal retrieval, i.e., recovering an accurate estimate of the input voltage given the waveform measured at the sensor output. The presence of a strong sensor resonance complicates this objective; the $>$30~dB variation in sensitivity between the flat and resonant portions of the sensor spectrum requires careful frequency matching spanning several orders of magnitude. Of course, one mitigation approach would be to simply filter out all spectral content beyond $\sim$1~kHz and restrict sensor attention to lower frequencies, yet this is undesirable, discarding potentially useful information contained within a significant portion of the sensor's response band. Accordingly, in the following we introduce a digital signal processing approach designed to recover the original input waveform up to the full bandwidth of the system.

\begin{figure*}[tb!]
\centering\includegraphics[width=5.85in]{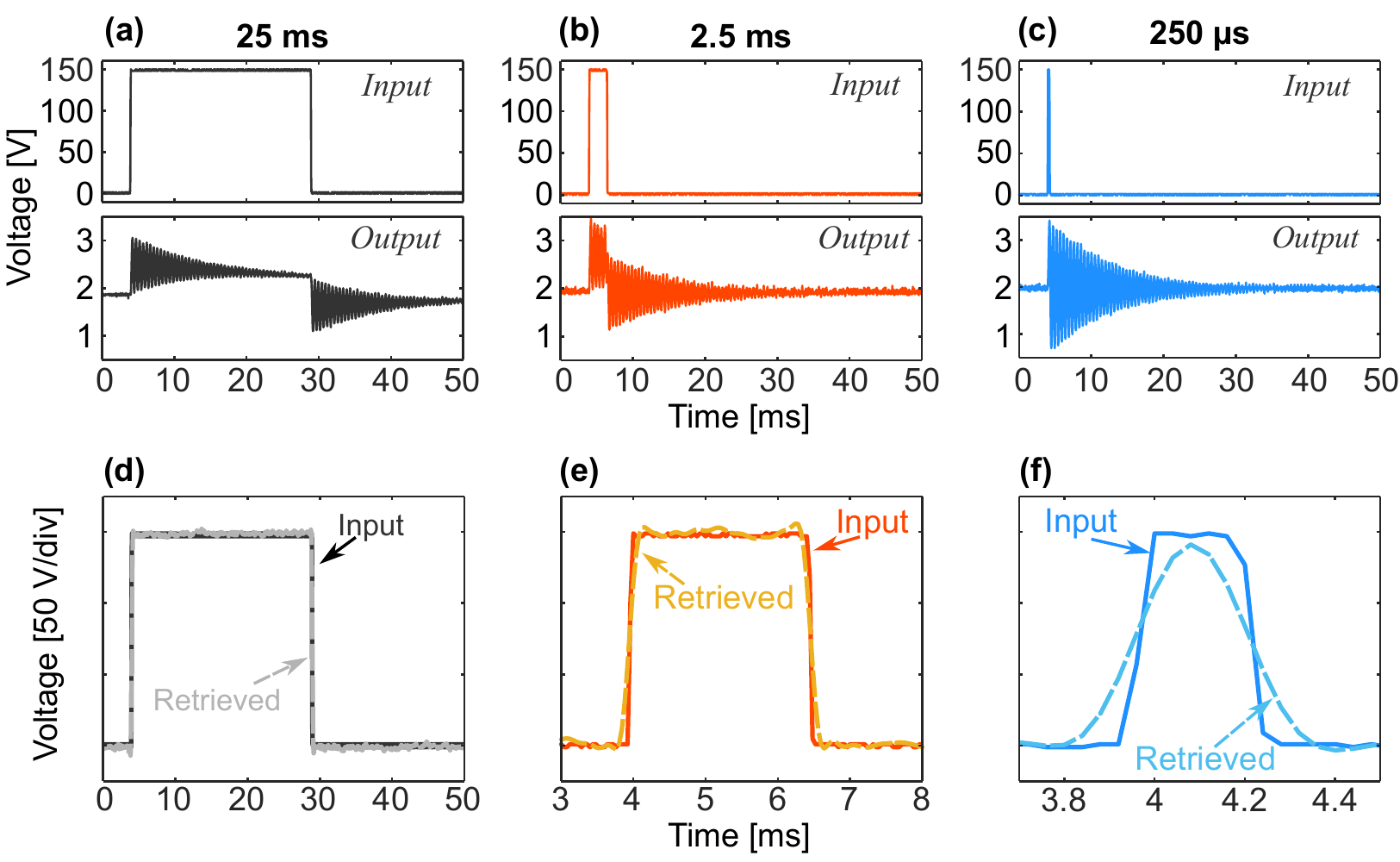}
\caption{Transient voltage tests. (a-c) Input and raw sensor output waveforms for square pulses of duration (a) 25~ms, (b) 2.5~ms, and (c) 250~$\upmu$s. (d-f) Digitally retrieved estimates of excitation pulses, compared to the actual inputs, for the cases (d) 25~ms, (e) 2.5~ms, and (f) 250~$\upmu$s.}
\label{fig6}
\end{figure*}

As example test cases, we probe each sensor with short electrical pulses containing frequency content well into the resonant regime. Figure~\ref{fig6} furnishes single-shot measurements for three examples of 150~V~peak square waves, all sampled at 25~kS/s, with a wide range of durations: (a) 25~ms, applied to the phase 1 sensor; (b) 2.5~ms, applied to phase 2; and (c) 250~$\upmu$s, applied to phase 3. In all cases, the raw output exhibits significant distortion, with strong 2~kHz ringing, and fails to accurately trace the shape of the input. (Note that the output offsets of $\sim$2~V are the quiescent points of the sensor; they do not reflect dc components in the input.) Interestingly, the oscillations are particularly strong for the 250~$\upmu$s excitation; because the positive and negative pulse edges are spaced by half a period at 2~kHz, their contributions add in-phase.

In practice, simply dividing the distorted spectrum of the output, $V_\mathrm{out} (\omega)$, by $H(\omega)$ introduces error at portions where the transfer function spectrum is low, artificially amplifying frequency components to which the sensor does not actually respond. Additionally, frequency-truncated digital reconstruction of jump discontinuities, such as those present here, experience overshoot from the well-known Gibbs phenomenon. To reduce the impact of both effects, we multiply $V_\mathrm{out}(\omega)$ by an apodization function before dividing out $H(\omega)$. Specifically, we consider the product of two sinusoidal windows, one for removing low-frequency content, the other for high, defined over positive frequencies as
\begin{equation}
W_L(\omega) = \begin{cases}
\sin\frac{\pi\omega}{2\omega_L} & 0<\omega<\omega_L \\
1 & \omega_L < \omega
\end{cases}
\end{equation}
and
\begin{equation}
W_H(\omega) = \begin{cases}
\cos\frac{\pi\omega}{2\omega_H} & 0<\omega<\omega_H \\
0 & \omega_H < \omega
\end{cases}.
\end{equation}
For the low and high frequencies, we choose $\omega_L/2\pi$ = 10~Hz and $\omega_H/2\pi = 4$~kHz. We emphasize that the specific functional forms of these windows represent just one convenient option; a variety of filter selections would achieve comparable apodization performance. Our procedure for returning an estimate $\tilde{v}_\mathrm{in} (t)$ of the input signal from the raw output $v_\mathrm{out} (t)$ amounts to computing the Fourier transform $V_\mathrm{out} (\omega)$ and calculating the estimated input spectrum via
\begin{equation}
\tilde{V}_\mathrm{in} (\omega) = \frac{W_L(\omega) W_H (\omega)}{H(\omega)} V_\mathrm{out}(\omega),
\end{equation}
from which $\tilde{v}_\mathrm{in}(t)$ follows from Fourier inversion.

The waveforms retrieved in this process are given in Fig.~\ref{fig6}(d), (e), and (f), corresponding to the 25~ms, 2.5~ms, and 250~$\upmu$s excitations, respectively. We plot on a relative voltage scale here, because any dc offset is not meaningful given the sensor response. The previous 2~kHz ringing has been removed, and the reconstructions show good agreement with the true input pulses, particularly in the 25 and 2.5~ms cases. For the most extreme example of 250~$\upmu$s, we begin to see the impact of bandwidth limitations, with the retrieved waveform noticeably widened relative to the input. Yet while the retrieved amplitude in (f) is $\sim$5\% away from the actual value, were we to filter out the resonant region entirely for simpler reconstruction---e.g., by setting $\omega_H/2\pi = 1$~kHz---the retrieved pulse height would fall short of the actual by a massive 68\%, highlighting the importance of including the resonance in reconstruction. Finally, since our method utilizes linear digital filtering with fixed and known functions, it could be implemented on a field-programmable gate array (FPGA) for spectral equalization of the sensor output in real-time. In this way the sharp physical resonance presents no fundamental limitation to accurate voltage sensing, and can even be leveraged in expanding sensor bandwidth.

\section{Discussion and Conclusion}

Making use of Lorentz-force-based displacement, the fiber-optic sensing approach employed here can be applied to monitoring electric current as well~\cite{Lagakos2017}. The basic idea is to shunt a portion of the current from the main conductor into a secondary wire. Then, by taking advantage of the main conductor's intrinsic magnetic field (or that of a dedicated permanent magnet), the shunting conductor will displace in proportion to the carried current, thereby providing the mechanical movement necessary for optical probe and readout. We plan to conduct additional inquiries into the use of this approach to measure current, including new experiments as well as analyzing data from past exercises. Moreover, while here we have concentrated on voltages up to 150~V~peak, which are relatively low from a power systems perspective, the same sensing technology is scalable to much higher, kilovolt distribution and transmission levels as well, where we anticipate similar physical behavior and thus applicability of our equalization method. We plan to examine in detail both current sensors and higher voltage levels in future work. It will prove valuable to perform tests of these intensity-modulated optical sensors in side-by-side comparisons against both conventional devices, such as potential transformers (PTs) and current transformers (CTs)~\cite{Horowitz2008}, and more complex interferometric-based optical phase/polarity sensors---particularly undertaking such comparisons in demanding field environments. Finally, we anticipate comparing sensor probe/transducer combinations with differing resonance points and frequency ranges.

Ultimately, a low-cost solution for the system should be feasible as well. The LED, fiber, and transducer elements are naturally inexpensive, so that total system cost is dominated by the electro-optic circuity. Consequently, further investment and high-volume production should bring the cost down significantly, which we expect will make the system economically competitive with traditional PTs. 

In conclusion, we have described and characterized intensity-modulated fiber-optic voltage sensors. Probing with a tunable-frequency source, we measure sensor spectra exemplified by a flat response from 10 to 500 Hz and a sharp resonance at 2~kHz. The measured background noise levels imply a full-band dynamic range of approximately 60~dB. And through a digital spectral equalization method, we have demonstrated successful reconstruction of short-pulse inputs from strongly distorted output waveforms. The simplicity and robustness of these intensity-modulated sensors offer a valuable balance between the advantage tradeoffs of fully electrical sensors (low cost) and their optical interferometric replacements (high performance), and could find application in a variety of electromagnetic sensing environments.

\section*{Funding}
U.S. Department of Energy, Office of Electricity (field work proposal CETE004).

\section*{Acknowledgments}
We are grateful to B. Qi for valuable discussions and feedback. A portion of this work was performed at Oak Ridge National Laboratory, operated by UT-Battelle for the U.S. Department of Energy under contract no. DE-AC05-00OR22725.

\newpage


\begin{thebibliography}{23}%
\makeatletter
\providecommand \@ifxundefined [1]{%
 \@ifx{#1\undefined}
}%
\providecommand \@ifnum [1]{%
 \ifnum #1\expandafter \@firstoftwo
 \else \expandafter \@secondoftwo
 \fi
}%
\providecommand \@ifx [1]{%
 \ifx #1\expandafter \@firstoftwo
 \else \expandafter \@secondoftwo
 \fi
}%
\providecommand \natexlab [1]{#1}%
\providecommand \enquote  [1]{``#1''}%
\providecommand \bibnamefont  [1]{#1}%
\providecommand \bibfnamefont [1]{#1}%
\providecommand \citenamefont [1]{#1}%
\providecommand \href@noop [0]{\@secondoftwo}%
\providecommand \href [0]{\begingroup \@sanitize@url \@href}%
\providecommand \@href[1]{\@@startlink{#1}\@@href}%
\providecommand \@@href[1]{\endgroup#1\@@endlink}%
\providecommand \@sanitize@url [0]{\catcode `\\12\catcode `\$12\catcode
  `\&12\catcode `\#12\catcode `\^12\catcode `\_12\catcode `\%12\relax}%
\providecommand \@@startlink[1]{}%
\providecommand \@@endlink[0]{}%
\providecommand \url  [0]{\begingroup\@sanitize@url \@url }%
\providecommand \@url [1]{\endgroup\@href {#1}{\urlprefix }}%
\providecommand \urlprefix  [0]{URL }%
\providecommand \Eprint [0]{\href }%
\providecommand \doibase [0]{http://dx.doi.org/}%
\providecommand \selectlanguage [0]{\@gobble}%
\providecommand \bibinfo  [0]{\@secondoftwo}%
\providecommand \bibfield  [0]{\@secondoftwo}%
\providecommand \translation [1]{[#1]}%
\providecommand \BibitemOpen [0]{}%
\providecommand \bibitemStop [0]{}%
\providecommand \bibitemNoStop [0]{.\EOS\space}%
\providecommand \EOS [0]{\spacefactor3000\relax}%
\providecommand \BibitemShut  [1]{\csname bibitem#1\endcsname}%
\let\auto@bib@innerbib\@empty
%</preamble>
\bibitem [{\citenamefont {Udd}\ and\ \citenamefont
  {Spillman~Jr}(2011)}]{Udd2011}%
  \BibitemOpen
  \bibinfo {editor} {\bibfnamefont {E.}~\bibnamefont {Udd}}\ and\ \bibinfo
  {editor} {\bibfnamefont {W.~B.}\ \bibnamefont {Spillman~Jr}},\ eds.,\
  \href@noop {} {\emph {\bibinfo {title} {Fiber Optic Sensors: An Introduction
  for Engineers and Scientists}}}\ (\bibinfo  {publisher} {Wiley},\ \bibinfo
  {address} {New York, NY},\ \bibinfo {year} {2011})\BibitemShut {NoStop}%
\bibitem [{\citenamefont {Krohn}\ \emph {et~al.}(2014)\citenamefont {Krohn},
  \citenamefont {MacDougall},\ and\ \citenamefont {Mendez}}]{Krohn2014}%
  \BibitemOpen
  \bibfield  {author} {\bibinfo {author} {\bibfnamefont {D.~A.}\ \bibnamefont
  {Krohn}}, \bibinfo {author} {\bibfnamefont {T.}~\bibnamefont {MacDougall}}, \
  and\ \bibinfo {author} {\bibfnamefont {A.}~\bibnamefont {Mendez}},\
  }\href@noop {} {\emph {\bibinfo {title} {Fiber Optic Sensors: Fundamentals
  and Applications}}},\ \bibinfo {edition} {4th}\ ed.\ (\bibinfo  {publisher}
  {SPIE Press},\ \bibinfo {address} {Bellingham, WA},\ \bibinfo {year}
  {2014})\BibitemShut {NoStop}%
\bibitem [{\citenamefont {Lee}(2003)}]{Lee2003}%
  \BibitemOpen
  \bibfield  {author} {\bibinfo {author} {\bibfnamefont {B.}~\bibnamefont
  {Lee}},\ }\href@noop {} {\bibfield  {journal} {\bibinfo  {journal} {Opt.
  Fiber Technol.}\ }\textbf {\bibinfo {volume} {9}},\ \bibinfo {pages} {57}
  (\bibinfo {year} {2003})}\BibitemShut {NoStop}%
\bibitem [{\citenamefont {{Culshaw}}\ and\ \citenamefont
  {{Kersey}}(2008)}]{Culshaw2008}%
  \BibitemOpen
  \bibfield  {author} {\bibinfo {author} {\bibfnamefont {B.}~\bibnamefont
  {{Culshaw}}}\ and\ \bibinfo {author} {\bibfnamefont {A.}~\bibnamefont
  {{Kersey}}},\ }\href@noop {} {\bibfield  {journal} {\bibinfo  {journal} {J.
  Light. Technol.}\ }\textbf {\bibinfo {volume} {26}},\ \bibinfo {pages} {1064}
  (\bibinfo {year} {2008})}\BibitemShut {NoStop}%
\bibitem [{\citenamefont {Lef\`{e}vre}(2014)}]{Lefevre2014}%
  \BibitemOpen
  \bibfield  {author} {\bibinfo {author} {\bibfnamefont {H.~C.}\ \bibnamefont
  {Lef\`{e}vre}},\ }\href@noop {} {\emph {\bibinfo {title} {The Fiber-Optic
  Gyroscope}}},\ \bibinfo {edition} {2nd}\ ed.\ (\bibinfo  {publisher} {Artech
  House},\ \bibinfo {address} {Boston, MA},\ \bibinfo {year}
  {2014})\BibitemShut {NoStop}%
\bibitem [{\citenamefont {Dandridge}(2019)}]{Dandridge2019}%
  \BibitemOpen
  \bibfield  {author} {\bibinfo {author} {\bibfnamefont {A.}~\bibnamefont
  {Dandridge}},\ }\href@noop {} {\bibfield  {journal} {\bibinfo  {journal}
  {Opt. Photon. News}\ }\textbf {\bibinfo {volume} {30(6)}},\ \bibinfo {pages}
  {34} (\bibinfo {year} {2019})}\BibitemShut {NoStop}%
\bibitem [{\citenamefont {{Cranch}}\ \emph {et~al.}(2003)\citenamefont
  {{Cranch}}, \citenamefont {{Nash}},\ and\ \citenamefont
  {{Kirkendall}}}]{Cranch2003}%
  \BibitemOpen
  \bibfield  {author} {\bibinfo {author} {\bibfnamefont {G.~A.}\ \bibnamefont
  {{Cranch}}}, \bibinfo {author} {\bibfnamefont {P.~J.}\ \bibnamefont
  {{Nash}}}, \ and\ \bibinfo {author} {\bibfnamefont {C.~K.}\ \bibnamefont
  {{Kirkendall}}},\ }\href {\doibase 10.1109/JSEN.2003.810102} {\bibfield
  {journal} {\bibinfo  {journal} {IEEE Sens. J.}\ }\textbf {\bibinfo {volume}
  {3}},\ \bibinfo {pages} {19} (\bibinfo {year} {2003})}\BibitemShut {NoStop}%
\bibitem [{\citenamefont {Ziegler}\ \emph {et~al.}(2009)\citenamefont
  {Ziegler}, \citenamefont {Woodward}, \citenamefont {Iu},\ and\ \citenamefont
  {Borle}}]{Ziegler2009}%
  \BibitemOpen
  \bibfield  {author} {\bibinfo {author} {\bibfnamefont {S.}~\bibnamefont
  {Ziegler}}, \bibinfo {author} {\bibfnamefont {R.~C.}\ \bibnamefont
  {Woodward}}, \bibinfo {author} {\bibfnamefont {H.~H.}\ \bibnamefont {Iu}}, \
  and\ \bibinfo {author} {\bibfnamefont {L.~J.}\ \bibnamefont {Borle}},\ }\href
  {\doibase 10.1109/JSEN.2009.2013914} {\bibfield  {journal} {\bibinfo
  {journal} {IEEE Sens. J.}\ }\textbf {\bibinfo {volume} {9}},\ \bibinfo
  {pages} {354} (\bibinfo {year} {2009})}\BibitemShut {NoStop}%
\bibitem [{\citenamefont {Massey}\ \emph {et~al.}(1975)\citenamefont {Massey},
  \citenamefont {Erickson},\ and\ \citenamefont {Kadlec}}]{Massey1975}%
  \BibitemOpen
  \bibfield  {author} {\bibinfo {author} {\bibfnamefont {G.~A.}\ \bibnamefont
  {Massey}}, \bibinfo {author} {\bibfnamefont {D.~C.}\ \bibnamefont
  {Erickson}}, \ and\ \bibinfo {author} {\bibfnamefont {R.~A.}\ \bibnamefont
  {Kadlec}},\ }\href@noop {} {\bibfield  {journal} {\bibinfo  {journal} {Appl.
  Opt.}\ }\textbf {\bibinfo {volume} {14}},\ \bibinfo {pages} {2712} (\bibinfo
  {year} {1975})}\BibitemShut {NoStop}%
\bibitem [{\citenamefont {Rashleigh}\ and\ \citenamefont
  {Ulrich}(1979)}]{Rashleigh1979}%
  \BibitemOpen
  \bibfield  {author} {\bibinfo {author} {\bibfnamefont {S.~C.}\ \bibnamefont
  {Rashleigh}}\ and\ \bibinfo {author} {\bibfnamefont {R.}~\bibnamefont
  {Ulrich}},\ }\href@noop {} {\bibfield  {journal} {\bibinfo  {journal} {Appl.
  Phys. Lett.}\ }\textbf {\bibinfo {volume} {34}},\ \bibinfo {pages} {768}
  (\bibinfo {year} {1979})}\BibitemShut {NoStop}%
\bibitem [{\citenamefont {Bohnert}\ \emph {et~al.}(2002)\citenamefont
  {Bohnert}, \citenamefont {Gabus}, \citenamefont {Nehring},\ and\
  \citenamefont {Brandle}}]{Bohnert2002}%
  \BibitemOpen
  \bibfield  {author} {\bibinfo {author} {\bibfnamefont {K.}~\bibnamefont
  {Bohnert}}, \bibinfo {author} {\bibfnamefont {P.}~\bibnamefont {Gabus}},
  \bibinfo {author} {\bibfnamefont {J.}~\bibnamefont {Nehring}}, \ and\
  \bibinfo {author} {\bibfnamefont {H.}~\bibnamefont {Brandle}},\ }\href
  {http://jlt.osa.org/abstract.cfm?URI=jlt-20-2-267} {\bibfield  {journal}
  {\bibinfo  {journal} {J. Lightwave Technol.}\ }\textbf {\bibinfo {volume}
  {20}},\ \bibinfo {pages} {267} (\bibinfo {year} {2002})}\BibitemShut
  {NoStop}%
\bibitem [{\citenamefont {Bohnert}\ \emph {et~al.}(2005)\citenamefont
  {Bohnert}, \citenamefont {Gabus}, \citenamefont {Kostovic},\ and\
  \citenamefont {Br\"{a}ndle}}]{Bohnert2005}%
  \BibitemOpen
  \bibfield  {author} {\bibinfo {author} {\bibfnamefont {K.}~\bibnamefont
  {Bohnert}}, \bibinfo {author} {\bibfnamefont {P.}~\bibnamefont {Gabus}},
  \bibinfo {author} {\bibfnamefont {J.}~\bibnamefont {Kostovic}}, \ and\
  \bibinfo {author} {\bibfnamefont {H.}~\bibnamefont {Br\"{a}ndle}},\
  }\href@noop {} {\bibfield  {journal} {\bibinfo  {journal} {Opt. Lasers Eng.}\
  }\textbf {\bibinfo {volume} {43}},\ \bibinfo {pages} {511} (\bibinfo {year}
  {2005})}\BibitemShut {NoStop}%
\bibitem [{\citenamefont {Kissinger}(1967)}]{Kissinger1967}%
  \BibitemOpen
  \bibfield  {author} {\bibinfo {author} {\bibfnamefont {C.~D.}\ \bibnamefont
  {Kissinger}},\ }\href@noop {} {\bibfield  {journal} {\bibinfo  {journal}
  {U.S. Patent 3,327,584}\ } (\bibinfo {year} {1967})}\BibitemShut {NoStop}%
\bibitem [{\citenamefont {Krohn}(1987)}]{Krohn1987}%
  \BibitemOpen
  \bibfield  {author} {\bibinfo {author} {\bibfnamefont {D.~A.}\ \bibnamefont
  {Krohn}},\ }\href@noop {} {\bibfield  {journal} {\bibinfo  {journal} {Proc.
  SPIE}\ }\textbf {\bibinfo {volume} {718}},\ \bibinfo {pages} {2} (\bibinfo
  {year} {1987})}\BibitemShut {NoStop}%
\bibitem [{\citenamefont {Bucaro}\ \emph {et~al.}(2005)\citenamefont {Bucaro},
  \citenamefont {Lagakos}, \citenamefont {Houston}, \citenamefont {Jarzynski},\
  and\ \citenamefont {Zalalutdinov}}]{Bucaro2005}%
  \BibitemOpen
  \bibfield  {author} {\bibinfo {author} {\bibfnamefont {J.~A.}\ \bibnamefont
  {Bucaro}}, \bibinfo {author} {\bibfnamefont {N.}~\bibnamefont {Lagakos}},
  \bibinfo {author} {\bibfnamefont {B.~H.}\ \bibnamefont {Houston}}, \bibinfo
  {author} {\bibfnamefont {J.}~\bibnamefont {Jarzynski}}, \ and\ \bibinfo
  {author} {\bibfnamefont {M.}~\bibnamefont {Zalalutdinov}},\ }\href@noop {}
  {\bibfield  {journal} {\bibinfo  {journal} {J. Acoust. Soc. Am.}\ }\textbf
  {\bibinfo {volume} {118}},\ \bibinfo {pages} {1406} (\bibinfo {year}
  {2005})}\BibitemShut {NoStop}%
\bibitem [{\citenamefont {Bucaro}\ \emph {et~al.}(2013)\citenamefont {Bucaro},
  \citenamefont {Lagakos}, \citenamefont {Houston}, \citenamefont {Dey},\ and\
  \citenamefont {Zalalutdinov}}]{Bucaro2013}%
  \BibitemOpen
  \bibfield  {author} {\bibinfo {author} {\bibfnamefont {J.~A.}\ \bibnamefont
  {Bucaro}}, \bibinfo {author} {\bibfnamefont {N.}~\bibnamefont {Lagakos}},
  \bibinfo {author} {\bibfnamefont {B.~H.}\ \bibnamefont {Houston}}, \bibinfo
  {author} {\bibfnamefont {S.}~\bibnamefont {Dey}}, \ and\ \bibinfo {author}
  {\bibfnamefont {M.}~\bibnamefont {Zalalutdinov}},\ }\href@noop {} {\bibfield
  {journal} {\bibinfo  {journal} {J. Acoust. Soc. Am.}\ }\textbf {\bibinfo
  {volume} {133}},\ \bibinfo {pages} {832} (\bibinfo {year}
  {2013})}\BibitemShut {NoStop}%
\bibitem [{\citenamefont {Lagakos}\ \emph {et~al.}(2017)\citenamefont
  {Lagakos}, \citenamefont {Kaybulkin}, \citenamefont {Hernandez},\ and\
  \citenamefont {Vizas}}]{Lagakos2017}%
  \BibitemOpen
  \bibfield  {author} {\bibinfo {author} {\bibfnamefont {N.}~\bibnamefont
  {Lagakos}}, \bibinfo {author} {\bibfnamefont {V.}~\bibnamefont {Kaybulkin}},
  \bibinfo {author} {\bibfnamefont {P.}~\bibnamefont {Hernandez}}, \ and\
  \bibinfo {author} {\bibfnamefont {C.}~\bibnamefont {Vizas}},\ }\href@noop {}
  {\bibfield  {journal} {\bibinfo  {journal} {U.S. Patent 9,823,277}\ }
  (\bibinfo {year} {2017})}\BibitemShut {NoStop}%
\bibitem [{\citenamefont {{U.S. Department of Defense}}(1995)}]{DoD1995}%
  \BibitemOpen
  \bibfield  {author} {\bibinfo {author} {\bibnamefont {{U.S. Department of
  Defense}}},\ }\href@noop {} {\bibfield  {journal} {\bibinfo  {journal}
  {MIL-STD-1376B(SH)}\ } (\bibinfo {year} {1995})}\BibitemShut {NoStop}%
\bibitem [{\citenamefont {Rumyantsev}\ \emph {et~al.}(2004)\citenamefont
  {Rumyantsev}, \citenamefont {Shur}, \citenamefont {Bilenko}, \citenamefont
  {Kosterin},\ and\ \citenamefont {Salzberg}}]{Rumyantsev2004}%
  \BibitemOpen
  \bibfield  {author} {\bibinfo {author} {\bibfnamefont {S.~L.}\ \bibnamefont
  {Rumyantsev}}, \bibinfo {author} {\bibfnamefont {M.~S.}\ \bibnamefont
  {Shur}}, \bibinfo {author} {\bibfnamefont {Y.}~\bibnamefont {Bilenko}},
  \bibinfo {author} {\bibfnamefont {P.~V.}\ \bibnamefont {Kosterin}}, \ and\
  \bibinfo {author} {\bibfnamefont {B.~M.}\ \bibnamefont {Salzberg}},\ }\href
  {\doibase 10.1063/1.1763225} {\bibfield  {journal} {\bibinfo  {journal} {J.
  Appl. Phys.}\ }\textbf {\bibinfo {volume} {96}},\ \bibinfo {pages} {966}
  (\bibinfo {year} {2004})}\BibitemShut {NoStop}%
\bibitem [{\citenamefont {Hooker}(1998)}]{Hooker1998}%
  \BibitemOpen
  \bibfield  {author} {\bibinfo {author} {\bibfnamefont {M.~W.}\ \bibnamefont
  {Hooker}},\ }\href@noop {} {\bibfield  {journal} {\bibinfo  {journal}
  {NASA/CR-1998-208708}\ } (\bibinfo {year} {1998})}\BibitemShut {NoStop}%
\bibitem [{\citenamefont {Siebert}(1986)}]{Siebert1986}%
  \BibitemOpen
  \bibfield  {author} {\bibinfo {author} {\bibfnamefont {W.~M.}\ \bibnamefont
  {Siebert}},\ }\href@noop {} {\emph {\bibinfo {title} {Circuits, Signals, and
  Systems}}}\ (\bibinfo  {publisher} {MIT Press},\ \bibinfo {address}
  {Cambridge, MA},\ \bibinfo {year} {1986})\BibitemShut {NoStop}%
\bibitem [{\citenamefont {Papoulis}\ and\ \citenamefont
  {Pillai}(2002)}]{Papoulis2002}%
  \BibitemOpen
  \bibfield  {author} {\bibinfo {author} {\bibfnamefont {A.}~\bibnamefont
  {Papoulis}}\ and\ \bibinfo {author} {\bibfnamefont {S.~U.}\ \bibnamefont
  {Pillai}},\ }\href@noop {} {\emph {\bibinfo {title} {Probability, Random
  Variables, and Stochastic Processes}}},\ \bibinfo {edition} {4th}\ ed.\
  (\bibinfo  {publisher} {McGraw-Hill},\ \bibinfo {address} {New York, NY},\
  \bibinfo {year} {2002})\BibitemShut {NoStop}%
\bibitem [{\citenamefont {Horowitz}\ and\ \citenamefont
  {Phadke}(2008)}]{Horowitz2008}%
  \BibitemOpen
  \bibfield  {author} {\bibinfo {author} {\bibfnamefont {S.~H.}\ \bibnamefont
  {Horowitz}}\ and\ \bibinfo {author} {\bibfnamefont {A.~G.}\ \bibnamefont
  {Phadke}},\ }\href@noop {} {\emph {\bibinfo {title} {Power System
  Relaying}}},\ \bibinfo {edition} {4th}\ ed.\ (\bibinfo  {publisher} {Wiley},\
  \bibinfo {address} {Hoboken, NJ},\ \bibinfo {year} {2008})\BibitemShut
  {NoStop}%
\end{thebibliography}
\end{document}